\documentclass{article}
\usepackage{graphicx, psfrag, epsf, amsmath, natbib, url}
\title{Multiple Systems Analysis for the quantification of Modern Slavery: Classical and Bayesian approaches}
\author{Bernard W. Silverman\\Rights Lab\\University of Nottingham\\ Nottingham, U.K.}

\begin{document}
\maketitle
\begin{abstract}
Multiple systems estimation is a key approach for quantifying hidden populations such as the number of victims of modern slavery. The UK Government published an estimate of 10,000 to 13,000 victims, constructed by the present author, as part of the strategy leading to the Modern Slavery Act 2015.  This estimate was obtained by a stepwise multiple systems method based on six lists. Further investigation shows that a small proportion of the possible models give rather different answers, and that other model fitting approaches may choose one of these. Three data sets collected in the field of modern slavery, together with a data set about the death toll in the Kosovo conflict, are used to investigate the stability and robustness of various multiple systems estimate methods. The crucial aspect is the way that interactions between lists are modelled, because these can substantially affect the results. Model selection and Bayesian approaches are considered in detail, in particular to assess their stability and robustness when applied to real modern slavery data. A new Markov Chain Monte Carlo Bayesian approach is developed; overall, this gives robust and stable results at least for the examples considered. The software and datasets are freely and publicly available to facilitate wider implementation and further research.
\end{abstract}

\hyphenation{modslavmse}
\section{Introduction}
The original motivation for this work came from the estimation of the number of `potential victims of human trafficking' in the UK, based on the National Crime Agency strategic assessment of 2013. This was part of the strategy leading to the Modern Slavery Act 2015. See \cite{Sil14} and \cite{BHS15}. The method used was multiple systems estimation. 

Quantifying modern slavery has crucial importance for policy.  For example \cite{Co15} has written `without good data on where slaves are, how they become slaves and what happens to them, anti-slavery policy will remain guesswork', and goes on in this context to cite the use of multiple systems approaches as a significant innovative approach in a field where good quantification is in its infancy.  It is not just in narrow policy terms that good prevalence estimates are important; they also play a vital role in raising the public and political consciousness of modern slavery.

Multiple systems estimation is a development of the classical capture-recapture approach and has been used in many different contexts, such
as counting casualties in armed conflicts \citep{MPG13, 2019arXiv190604763M} and numbers of injecting drug users \citep{KBOHH13}.   Cases that come to light are recorded on a number of different lists.  By identifying cases across the various lists, the numbers that fall on each possible combination of lists are tabulated.  Then a mathematical model is used to estimate the `dark figure' of cases that have not come to attention, and so are not recorded on any list.   For an overall survey, see \cite{Bird2018}.

Crucial to this approach is the choice of model, in particular deciding which interactions or correlations to allow between the various lists.  Some methods choose a particular model, while others seek a model averaging approach.   This paper reviews a number of methods and investigates their performance on a range of real data sets.  There is a deliberate focus on data collected in the area of modern slavery and human trafficking, because the primary aim of this paper is to develop methodology relevant to that area.  In addition one of the data sets considered, drawn from the wider human rights area, relates to deaths in the 
Kosovo conflict in 1999.  The choice of existing methods for discussion and review is again guided by our particular context, focusing on methods already proposed for the multiple systems analysis of human rights and modern slavery data.    

The modern slavery context presents particular challenges for the use of multiple systems analysis.   There is no `ground truth' available to investigate the accuracy of any estimates, and so we need to assess other properties of estimation methods.  For example, it is clearly desirable to have reasonable stability under operations such as combining or omitting lists with small counts or adjusting model parameters.   Also, if multiple systems estimation is to be used more widely to quantify modern slavery, it is important to consider the performance of the various possible approaches specifically on data sets of the kinds likely to be observed.  Furthermore it may be important that there should be an agreed standard approach, at least as a starting point for more detailed investigation, and it is hoped that our detailed comparative study may contribute to that.    

Another issue that has to be borne in mind is the extremely sensitive nature of the data.  Typically, much as one would like more details, such as covariate information, about the individuals observed in the study, these are not available to the statistical analyst.  Without giving assurances of confidentiality to individual victims, for example, it would often not be ethical or even possible to collect their data.   Collation of data between lists naturally involves sharing or matching information, but this is often done by a trusted individual who cannot reveal any details.   Indeed, on some occasions all details of the lists themselves, and even of the type of organisation that provided particular lists, have to be obfuscated.   

Our comparative study using real data sets and the methods so far proposed will demonstrate that, unfortunately, all the existing methods display instabilities of various kinds, sometimes dramatic, when tested on the real data sets. To address this issue, we introduce a Bayesian/thresholding approach that places prior distributions on the individual terms in the standard model.  

In Section \ref{sec:datasets} below, the various data sets are reviewed and tabulated.  Section~\ref{sec:model} sets out the standard Poisson model which underlies various possible approaches.  Section~\ref{subsec:modselapp} then examines frequentist approaches to model selection, including the one used by \cite{Sil14}. Two other, rather different, Bayesian methods have been proposed and these are investigated in Section~\ref{sec:bayes2}. In Section~\ref{sec:bayes1} our proposed Bayesian/thresholding method for the Poisson model is introduced.  This casts the problem in a form where a standard Markov Chain Monte Carlo package can be used to estimate the parameters, but there are some mathematical aspects that have to be taken into account for this to work.  The method is demonstrated on the various data sets; it appears to avoid some of the gross instabilities that can arise with the existing methods, but still requires care in its application.      Finally, some conclusions are drawn in Section~\ref{sec:conclusions}. 

A key factor in developing a standard approach is the open accessibility of data and of methodology.  All the data sets, together with R software to implement the methodology described in this paper, and to reproduce its results, are given in \cite{Sil18R}.   For some additional remarks about the importance of open data and open research, see \cite{Sil2018Delta}.
\section{The data sets}
\label{sec:datasets}

The full data analysed by \cite{Sil14}, broken down into six lists, are given in Table~\ref{tab:UKdat}.  The abbreviations of the lists are explained as follows:

\begin{tabular}{||l|l||}
\hline\hline
LA & Local authorities \\
NG & Non-government organisations such as charities \\
PF  & Police forces \\
GO & Government organisations, such as the Border Force \\
& \mbox{  } and the Gangmasters and Labour Abuse Authority \\
GP & The general public, through various routes \\
NCA &  The National Crime Agency \\ 
\hline \hline
\end{tabular}
\\

\begin{table}
\caption{\label{tab:UKdat}Potential victims of trafficking in the UK, 2013. Numbers of cases on each possible combination of lists.  For example there are 54 cases that appear only on the LA list, and 15 cases that appear on the overlap between LA and NG, but not on any others.  There is one case that appears on all four of LA, NG, PF and GO but not on the other two.  Those combinations of lists for which no cases were observed are omitted from the table, but are still taken into account in the analysis. From \cite{BHS15}.}
\centering
%
\begin{tabular}{|| p{6pt}p{6pt}p{6pt}p{6pt}p{6pt}c|r|| c || *{5}{p{6pt}}c |r ||}
\cline{1-7} \cline{9-15}
LA&NG&PF&GO&GP&NCA&count & &LA&NG&PF&GO&GP&NCA&count \\
\cline{1-7} \cline{9-15}
$\times$&&&&&&54  &  & &&$\times$&$\times$&&&69  \\  
&$\times$&&&&&463   &  &  &&$\times$&&$\times$&&10\\
&&$\times$&&&&907   &  &  &&$\times$&&&$\times$&31\\
&&&$\times$&&&695  &  &  &&&$\times$&$\times$&&8\\
&&&&$\times$&&316  &  &  &&&$\times$&&$\times$&6\\
&&&&&$\times$&57  &  &  &&&&$\times$&$\times$&1\\
\cline{1-7}\cline{9-15}
$\times$&$\times$&&&&&15  &  &  $\times$&$\times$&$\times$&&&&1\\
$\times$&&$\times$&&&&19  &  &  $\times$&$\times$&&$\times$&&&1\\
$\times$&&&$\times$&&&3  &  &  &$\times$&$\times$&$\times$&&&4\\
&$\times$&$\times$&&&&56  &  &  &$\times$&$\times$&&&$\times$&3\\
&$\times$&&$\times$&&&19  &  &  &&$\times$&$\times$&&$\times$&1\\
\cline{9-15}
&$\times$&&&$\times$&&1  &  &  $\times$&$\times$&$\times$&$\times$&&&1\\
&$\times$&&&&$\times$&3  &  &  &&&&&&\\
\cline{1-7}\cline{9-15}
\end{tabular} 
\end{table}

Some of the methods we consider do not deal with more than five lists, and so for some purposes we will combine the list PF with the list NCA to construct the `UK five-list' data set.  The NCA is not, strictly speaking, a police organisation, but it has many powers and characteristics in common with police forces and so combining these two lists is the natural way to reduce to a smaller number.  

In addition, the list GP raises issues because cases on this list may not always be specified in sufficient detail to allow for reliable matching with other lists.  Therefore, at least to test for the robustness of any results, it will be helpful to consider, in addition to the full and five-list data sets, a `UK four-list' data set constructed by omitting the GP list and combining the PF and NCA lists.  The total number of observed cases is 2744 for the five- and six-list data, but only 2428 for the four-list data set.  

A second important data set \citep{DiCrHeKr17,CrDiHe17} comprises six lists for identified victims in the Netherlands for the period  2010--2015.  
The data are given in Table \ref{tab:Ned}.  For a five-list version of these data, we combine the two smallest lists I and O.   The total number of observed cases in this data set is 8234.  

\begin{table}
\caption{\label{tab:Ned}Victims of trafficking in the Netherlands. Numbers of cases on each possible combination of lists, leaving out combinations for which no cases were observed.  The lists are as follows: P = National Police; K = Border Police; I = Inspectorate SZW (Ministry of Social Affairs and Employment); R = regional coordinators; O = residential treatment centres and shelters; Z = others (for example, ambulatory care centres, organizations providing legal services, 
Immigration and Naturalization Service).  Constructed from \cite{DiCrHeKr17}, Table 3.}
\begin{tabular}{||cccccc|r||c ||cccccc|r|| }
\cline{1-7} \cline{9-15}
I&K&O&P&R&Z&count& & I&K&O&P&R&Z&count \\
\cline{1-7} \cline{9-15}
$\times$&&&&&&352&   &	&$\times$&&&&$\times$&4\\
&$\times$&&&&&1299&   &&&$\times$&$\times$&&&59\\
&&$\times$&&&&403&   &&&$\times$&&$\times$&&2\\
&&&$\times$&&&4466&   &&&$\times$&&&$\times$&57\\
&&&&$\times$&&650&   &&&&$\times$&$\times$&&82\\
&&&&&$\times$&632&   & &&&$\times$&&$\times$&125\\
\cline{1-7}
$\times$&&$\times$&&&&1&                      & &&&&$\times$&$\times$&2\\
 
\cline{9-15}
$\times$&&&$\times$&&&18&   &$\times$&&$\times$&$\times$&&&4\\

$\times$&&&&$\times$&&3&   &$\times$&&&$\times$&&$\times$&4\\
$\times$&&&&&$\times$&16&   &&&$\times$&$\times$&$\times$&&2\\
&$\times$&$\times$&&&&1&   &&&$\times$&$\times$&&$\times$&7\\
&$\times$&&$\times$&&&44&   & &&&$\times$&$\times$&$\times$&1\\
\cline{1-7} \cline{9-15}\cline{1-7} \cline{9-15}

\end{tabular}

\end{table}

The third example is constructed from data collected by eight agencies in the New Orleans-Metairie Metropolitan Statistical Area (Greater New Orleans) and analysed by \cite{BaMuSi18}.  These include 185 individuals who interacted with law enforcement and service providers in Greater New Orleans during the year 2016.  They are given in Table \ref{tab:NewOrl}.     The sensitivity among the various agencies, partly for legal reasons, means that it is not possible even to label the lists themselves informatively.   No further information was available to the statistical analysis than the table itself, with lists labelled A to H.  Where it is necessary to reduce the number of lists, a five-list data set is constructed by combining the lists with the four smallest counts into a single list BEFG.

\begin{table}
\caption{\label{tab:NewOrl}Victims related to modern slavery and trafficking in New Orleans. Numbers of cases on each possible combination of lists, leaving out combinations for which no cases were observed.  
For reasons of confidentiality the lists are labelled uninformatively. From \cite{BaMuSi18}}
\centering

\begin{tabular}{||*{7}{p{2pt}}c|r|| c ||*{7}{p{2pt}}c|r||}
\cline{1-9} \cline{11-19}
A&B&C&D&E&F&G&H&count     &   &   A&B&C&D&E&F&G&H&count    \\
\cline{1-9} \cline{11-19}
$\times$&&&&&&&&25&           &     $\times$&&&&$\times$&&&&1\\
&$\times$&&&&&&&5&             &    &$\times$&&&&$\times$&&&1\\
&&$\times$&&&&&&70&				&		&&$\times$&$\times$&&&&&1\\
&&&$\times$&&&&&33&				&			&&$\times$&&$\times$&&&&1\\
&&&&$\times$&&&&6&					&		&&$\times$&&&&$\times$&&1\\
&&&&&$\times$&&&6&					&				&&&$\times$&$\times$&&&&2\\
&&&&&&$\times$&&6&					&				&&&&$\times$&&&$\times$&1\\
&&&&&&&$\times$&21&				&    &&&&&&&&\\
\cline{1-9} \cline{11-19}
$\times$&&$\times$&&&&&&1&  &   $\times$&&$\times$&&&&$\times$&&1\\
$\times$&&&$\times$&&&&&2&   &  $\times$&&&$\times$&$\times$&&&&1\\
\cline{1-9} \cline{11-19}

\end{tabular}

\end{table}

Finally, we consider a data set from a different area of human rights, that of determining the numbers of victims of armed conflict.  The data, due to \cite{Ball02}, relate to the numbers of those killed in Kosovo in a three month period in 1999.  They are available within the R package {\bf LCMCR} \citep{Man17R} and are reproduced in Table \ref{tab:Kosovo}.  This four-list data set, which includes 4400 known victims, displays high correlation between lists, and has larger numbers in the higher order three-list and four-list overlaps than do the modern slavery examples.   This is in the nature of the particular application and is highly unlikely to occur in any modern slavery data set.    

\begin{table}
\caption{\label{tab:Kosovo}Killings in the Kosovo war from March 20 to June 22, 1999, grouped into four lists. All 15 observable combinations have a non-zero count. 
EXH = exhumations; ABA = American Bar Association Central and East European Law Initiative; OSCE = Organization for Security and Cooperation in Europe; HRW = Human Rights Watch.
From  \cite{Man17R}.}
\centering

\begin{tabular}{||*{4}{p{18pt}}|r||c||*{4}{p{18pt}}|r|| }
\cline{1-5}\cline{7-11}
EXH & ABA & OSCE & HRW & Count &&  EXH & ABA & OSCE & HRW & Count\\ \cline{1-5}\cline{7-11}
$\times$  &  &  &  & 1131 && & $\times$  &  & $\times$  & 31 \\ 
& $\times$  &  &  & 845 && &  & $\times$  & $\times$  & 123 \\
\cline{7-11}
&  & $\times$  &  & 936 &&$\times$  & $\times$  & $\times$  &  & 181 \\
 &  &  & $\times$  & 306 && $\times$  & $\times$  &  & $\times$  & 18 \\
 \cline{1-5}  
$\times$  & $\times$  &  &  & 177 &&$\times$  &  & $\times$  & $\times$  & 42 \\ 
$\times$  &  & $\times$  &  & 228 &&& $\times$  & $\times$  & $\times$  & 32 \\ 
\cline{7-11}
$\times$  &  &  & $\times$  & 106 && $\times$  & $\times$  & $\times$  & $\times$  & 27 \\ 
& $\times$  & $\times$  &  & 217 &&  &&&&\\
\cline{1-5}\cline{7-11}

\end{tabular}

\end{table}

\section{Models and methods}
\label{sec:model}

In this section, we review the basic log-linear model as proposed by \cite{Cormack1989}.  
Suppose we have $K$ lists labelled $\{ 1 , 2, \ldots, K \}$.   For each subset $A$ of $\{ 1 , 2, \ldots, K \}$, let $N_A$ be the number of cases that occur on all the lists in $A$ but on no others. 
So, if $K=6$ there are 64 possible subsets $A$, including the empty set $\emptyset$.  
The `dark figure' is the number of cases $N_\emptyset$ that do not appear on any list. 

Using the UK data as an illustrative example, Table \ref{tab:UKdat} only gives counts for 26 subsets $A$, and the first step in the analysis is to reinstate all the rows in the table for which the observed count is zero, yielding 63 observations in all.  There is no observed count for the dark figure.  

The basic model is that each $N_A$ has, independently, a Poisson distribution with parameter $\lambda_A$, with some structure on the $\lambda_A$.  This is quite a strong assumption, because it assumes that the cases each behave independently of one another and obey the same probability laws of appearing on the various lists.  Especially if there are observed covariates, the model will only be a jumping off point for more detailed modelling, but it is at least a start.
The model does not assume that the various lists are independent; interactions between the lists are allowed by appropriate modelling of the parameters $\lambda_A$.  

Under the model, the dark figure $N_\emptyset \sim \rm{Poiss}(\lambda_\emptyset)$.  It is likely that the estimation error in $\lambda_\emptyset$ will be much larger than the Poisson variation, and so in practice the parameter estimate of $\lambda_\emptyset$ will be taken as the estimate of the dark figure, though if possible the Poisson variation should be taken into account too.  To get an estimate of the total population, the estimate of the dark figure is added to the total number of cases actually observed.  

The Poisson model can also be seen as an approximation to a multinomial model where there is a fixed (unknown) total population size, and cases independently fall on the various lists or combinations of lists with probabilities proportional to the expected values under the Poisson model.   \cite{Cormack1992} provides a way of using the profile likelihood under the Poisson model set out below, to obtain confidence intervals for the total population size under the multinomial model.  

\label{subsec:poisspar}

For the most part, the model we will investigate will be of the form
\begin{equation}
\label{eq:poisspar}
\log{\lambda_A} = \mu + \sum_{i \in A}\alpha_i +  \sum_{\substack{ i, j \in A\\ i<j}} \beta_{ij} 
\end{equation}
For example, if $K=6$ then there will be six main effects $\alpha_i$ and 15 two-list interactions $\beta_{ij}$, making 22 parameters altogether to be estimated from the 63 observable values $N_A$.  
Within this model, we have $\log{\lambda_\emptyset} = \mu$.  Therefore the estimate of the dark figure is $\exp{\mu}$; we do not actually need estimates of the other parameters to estimate the dark figure.

There are basically two approaches to model fitting in this context.  One is to use a model selection criterion to choose a particular set of parameters to fit, constraining all the others to zero.  The other is to use some sort of model averaging approach, usually of a Bayesian nature.

\section{Frequentist model selection}
\label{subsec:modselapp}

The package {\bf Rcapture} \citep{BR07} can be used as the basis of various approaches, which are explored in this section.  The simplest is to set all interaction terms to zero, fitting main effects $\alpha_i$ only. Under this model, the lists themselves are independent, an assumption that may be unrealistic.  Nevertheless this model may be a good reference point for more detailed analysis.  

\subsection{Adding parameters stepwise}

In their original work on the UK data, \cite{Sil14} and \cite{BHS15} used a stepwise approach, starting with main effects only and then adding two-list interactions $\beta_{ij}$ stepwise.  At each step, the interaction that best improves 
the AIC criterion is chosen.  The process of adding interactions is stopped if the AIC cannot be improved by adding an interaction, or if the new interaction is not significant at some threshold.  This variable selection method is implemented within the R package {\bf modslavmse} \citep{Sil18R}, and makes use
of the package {\bf Rcapture}.

Table \ref{tab:UKdatres1} shows estimates and confidence limits using main effects only, and the stepwise procedure with two different $p$-value thresholds, for the UK data summarised into six, five and four lists as set out in Section~\ref{sec:datasets}.  The original work used the stepwise method with $p = 5\%$.  
Here and subsequently in this section, the confidence intervals are constructed from the profile likelihood using the approach of \cite{Cormack1992} as implemented within {\bf Rcapture}.
Both 
the six- and five-list data give a 95\% confidence interval, conditional on the model choice, of 10,000 to 13,000 in round terms.   Using main effects only, or a more stringent criterion for adding parameters to the model, gives larger estimates.   The results for the four-list case, on the other hand, give smaller estimates, but none of these effects is dramatic.  

\begin{table}
\caption{Estimates and confidence intervals for the UK data, for the main effects model and for the stepwise AIC approach. The figures are for the numbers of thousands of victims, rounded to the nearest one hundred. The bold-face row corresponds
to the analysis carried out by \cite{Sil14}.\label{tab:UKdatres1}}

\centering

\begin{tabular}{||l|rrrrr|| }
\hline\hline
&\multicolumn{5}{c||}{Estimates and confidence limits}\\				
Data& 2.5\% & 10\% & point est & 90\% & 97.5\% \\ \hline
\multicolumn{6}{||c||}{Main effects only} \\ \hline
UK six lists & 11.0 & 11.4 & 12.2 & 13.1 & 13.6 \\
UK five lists & 12.0 & 12.5 & 13.4 & 14.5 & 15.2 \\
UK four lists & 9.5 & 9.9 & 10.7 & 11.6 & 12.1 \\ \hline \hline
\multicolumn{6}{||c||}{Stepwise AIC, threshold $p$-value 0.1\%} \\ \hline
UK six lists & 12.6 & 13.1 & 14.2 & 15.4 & 16.1 \\
UK five lists & 12.6 & 13.1 & 14.2 & 15.4 & 16.1 \\
UK four lists & 10.5 & 11.0 & 12.0 & 13.1 & 13.8 \\
 \hline \hline
\multicolumn{6}{||c||}{Stepwise AIC, threshold $p$-value 5\%} \\ \hline 
UK six lists & 10.0 & 10.4 & 11.4 & 12.5 & 13.2 \\
{\bf UK five lists} & {\bf 9.9} & {\bf 10.3} & {\bf 11.3} & {\bf 12.4} & {\bf 13.1}\\
UK four lists & 9.6 & 10.0 & 11.0 & 12.1 & 12.8 \\
\hline\hline
\end{tabular}

\end{table}

\subsection{Choosing from a large class of models using an information criterion}

The stepwise method is not the only possibility.  Another approach is to fit all  possible models, considering every subset of the interactions, and to choose between these using some criterion.  This can be done using the routine {\tt closedpMS.t} within the package {\bf Rcapture}.  
If the full six-list data are used, then there are $2^{15}$ models even if only pairwise interactions are considered, which presents an excessive computational burden.  To make the method computationally feasible in practice, the approach is only applied to the five-list data, allowing for two-list interactions only, leaving $ 2^{10}$  models to be considered.  
The package {\bf Rcapture} displays the results using the BIC rather than AIC as the primary method of model choice.  The BIC and AIC differ in the amount they correct for parsimony of models, with BIC having a heavier preference for more parsimonious models. 

\begin{figure}
\caption{Estimates of abundance plotted against BIC, with outliers omitted, the default plot option}
  \label{fig:BIComitoutliers}
\centering
 \includegraphics[height=8cm]{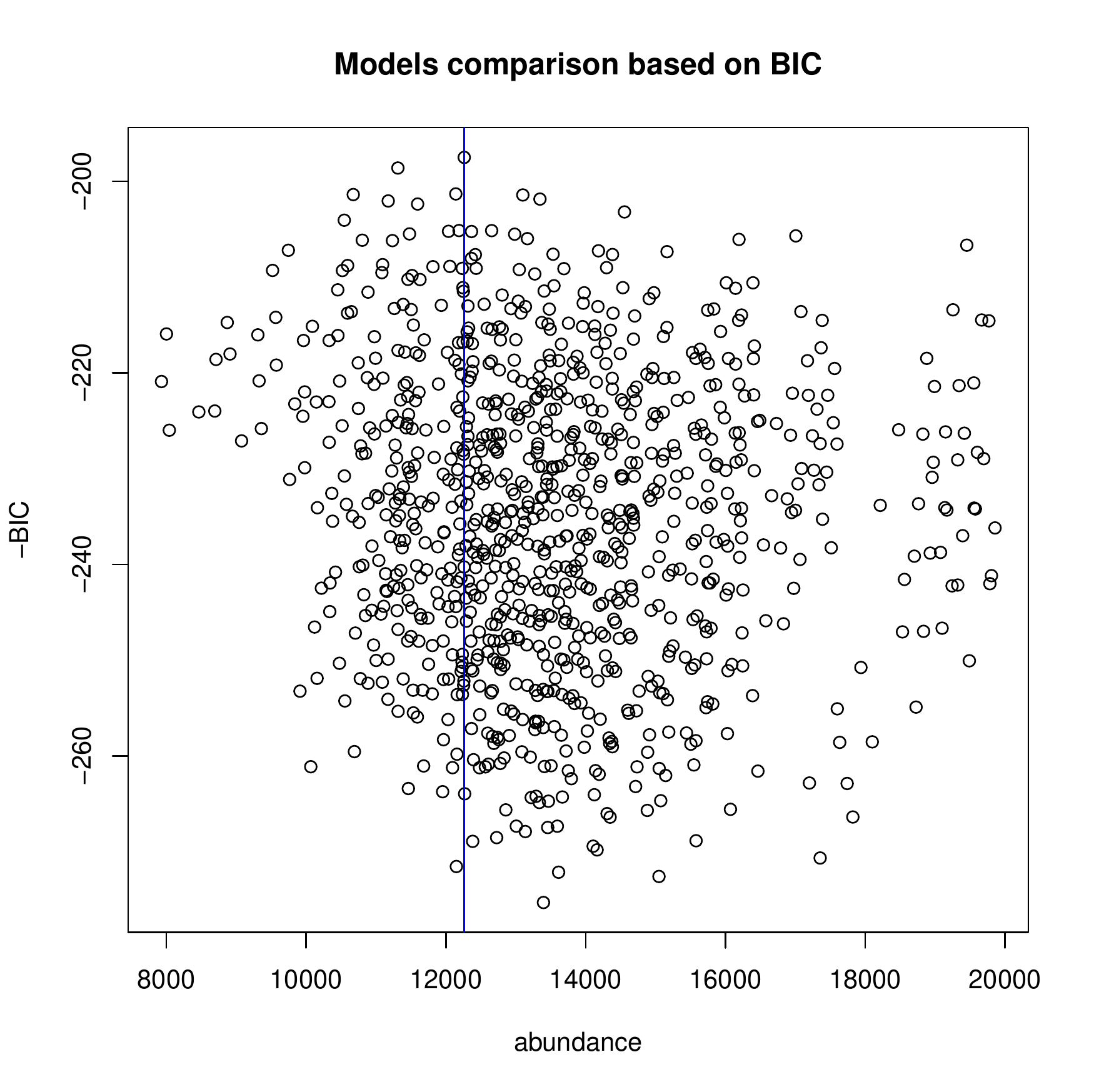}
  
\end{figure}
\begin{figure}
\caption{Estimates of abundance plotted against BIC, with outliers included}
  \label{fig:BICkeepoutliers}
\centering
 \includegraphics[height=8cm]{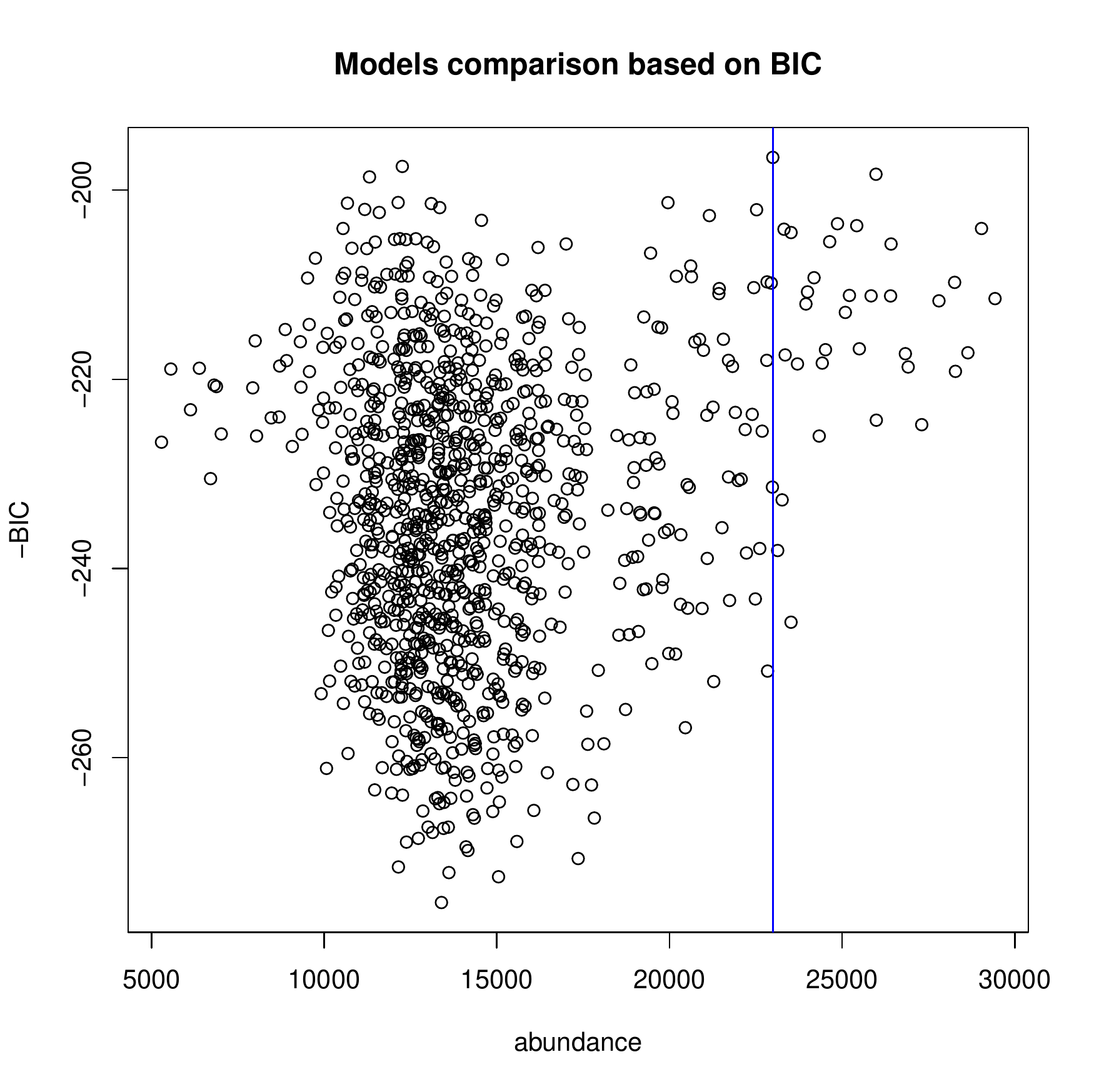}
  
\end{figure}
\begin{figure}
\caption{Estimates of abundance plotted against BIC, with list GP excluded}
  \label{fig:allmodelsomitGP}
\centering
  \includegraphics[height=8cm]{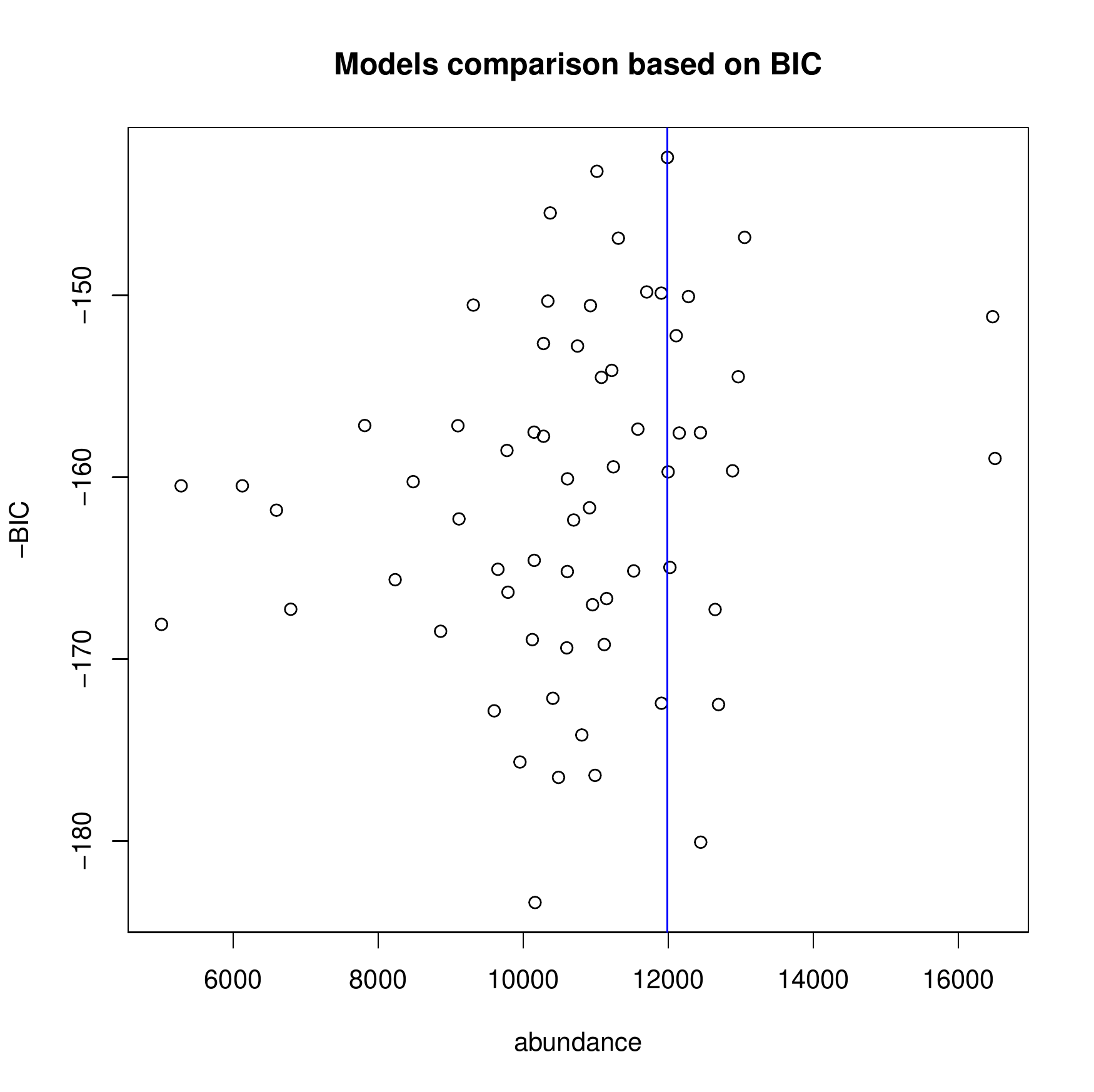}
  
\end{figure}

The default plot is shown in Figure \ref{fig:BIComitoutliers}, with the vertical line  showing the model with the lowest BIC (197.5). (The {\bf Rcapture} routine plots $-$BIC and chooses the maximum of that.) The population size estimate for that model is 12262, with the estimates for other models clustering approximately around the \cite{Sil14} estimate.  However setting the argument {\tt omitOutliers = F} yields Figure \ref{fig:BICkeepoutliers}.  There is a subsidiary cloud of results corresponding to a much larger estimate for the population size, and the estimate for the best BIC is actually within that cloud. 
 
Closer examination of the top ten models chosen by each of BIC and AIC is instructive.  There are no models in the top ten for AIC which yield estimates over 17,000, and only one which yields an estimate much outside the range suggested in the original analysis.  On the other hand, BIC chooses models yielding a much wider range of estimates.   Overall, the results for the five-list data demonstrate that the estimate of the total population can vary considerably depending on the model that is chosen, and that even concentrating on well-fitting models, by some criterion, does not necessarily resolve this issue. 

Because of the caveats about the GP list, the analysis was repeated for the four-list data. Figure \ref{fig:allmodelsomitGP} shows all models and demonstrates that the cloud of points corresponding to the much larger estimate disappears altogether if GP is omitted.  

\subsection{Further examples}

\begin{table}
\caption{Estimates and confidence intervals for the Netherlands, New Orleans and Kosovo data. The figures are for the numbers of thousands of victims, rounded to the nearest one hundred.
\label{tab:otherdatres1}}
\centering

\begin{tabular}{||l|rrrrr|| }
\hline\hline
&\multicolumn{5}{c||}{Estimates and confidence limits}\\				
Data& 2.5\% & 10\% & point est & 90\% & 97.5\% \\ \hline
\multicolumn{6}{||c||}{Main effects only} \\ \hline
Netherlands & 48.5 & 50.0 & 52.8 & 55.9 & 57.6 \\Netherlands five lists & 48.6 & 50.0 & 52.9 & 56.0 & 57.8 \\ \hline
 New Orleans & 0.7 & 0.7 & 1.0 & 1.4 & 1.7 \\New Orleans five lists & 0.7 & 0.8 & 1.0 & 1.4 & 1.8\\ \hline
Kosovo & 
7.1 & 7.2 & 7.4& 7.6& 7.7 \\ \hline \hline
\multicolumn{6}{||c||}{Stepwise AIC, threshold $p$-value 0.1\%} \\ \hline
Netherlands & 53.3 & 55.6 & 60.3 & 65.6 & 68.7 \\
Netherlands five lists & 119.4 & 127.8 & 146.0 & 167.8 & 181.0 \\ \hline
New Orleans & 0.7 & 0.7 & 1.0 & 1.4 & 1.7 \\
New Orleans five lists & 0.7 & 0.8 & 1.0 & 1.4 & 1.8 \\ \hline
Kosovo & 12.5 & 13.1 & 14.3 & 15.7 & 16.5 \\
\hline \hline
\multicolumn{6}{||c||}{Stepwise AIC, threshold $p$-value 5\%} \\ \hline 
Netherlands & 53.3 & 55.6 & 60.3 & 65.6 & 68.7 \\
Netherlands five lists & 119.4 & 127.8 & 146.0 & 167.8 & 181.0 \\ \hline
New Orleans & 1.4 & 1.8 & 3.4 & 7.2 & $\infty$ \\
New Orleans five lists & 0.7 & 0.8 & 1.0 & 1.4 & 1.8 \\ \hline
Kosovo & 12.5 & 13.1 & 14.3 & 15.7 & 16.5 \\
\hline\hline
\end{tabular}

\end{table}

\begin{figure}
\centering
  \includegraphics[width=15cm]{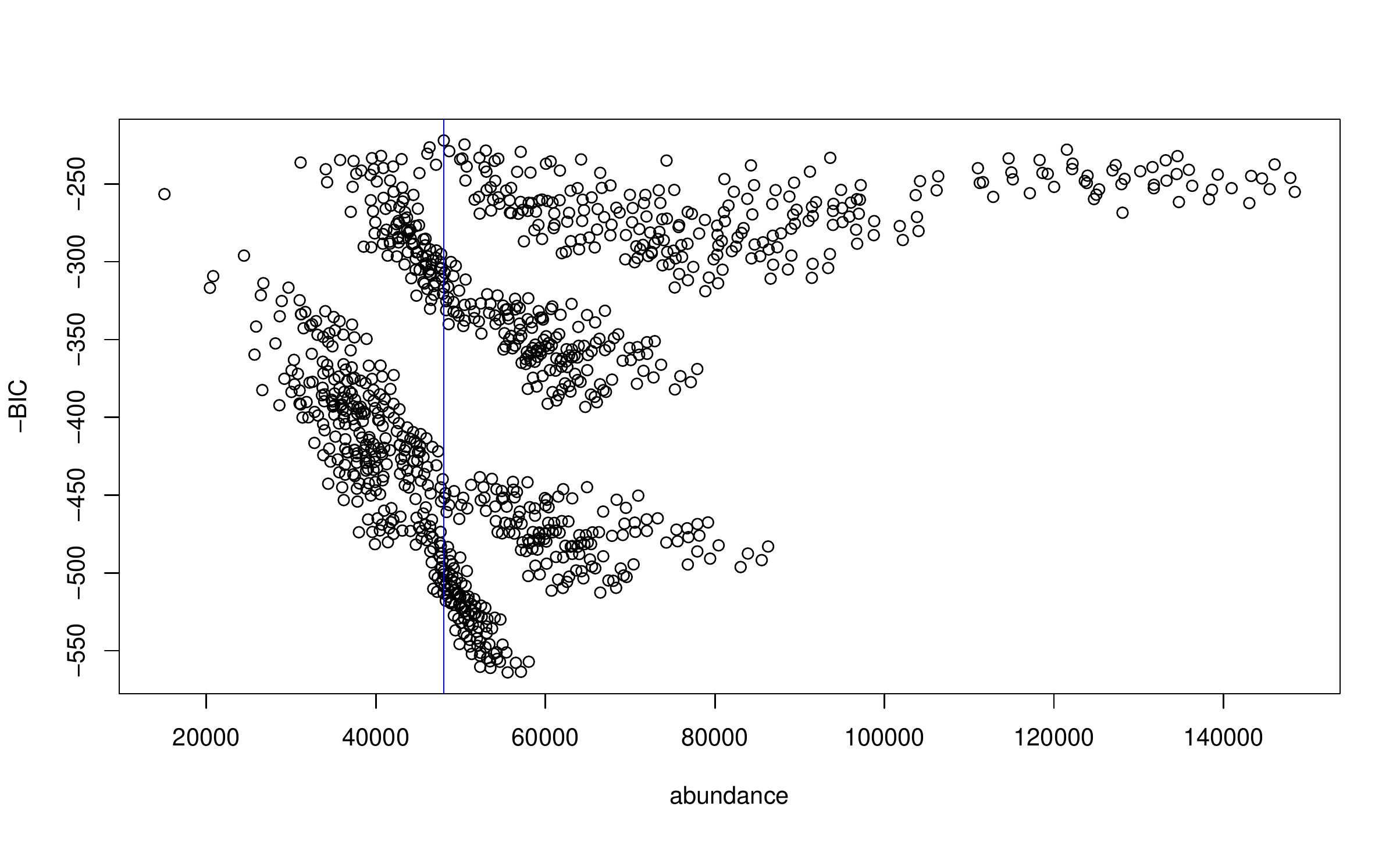}
  \caption{Estimates of abundance plotted against BIC, for the Netherlands data consolidated into five lists}
  \label{fig:Ned1AllModels}
\end{figure}

In Table \ref{tab:otherdatres1} we present the results of applying the main-effects only and the stepwise AIC choice methods to the other three example data sets.  
There is a somewhat alarming instability in the analysis of the Netherlands data; combining the two smallest lists more than doubles the stepwise estimates. There is no such instability if main effects only are fitted.   Some further intuition may be gained from Figure \ref{fig:Ned1AllModels}.  There is a long tail of models with very large estimates.  Although the globally optimal model according to BIC is not in this group, the stepwise method is choosing one of these, indeed one yielding almost the largest estimate among all choices of model.   

For the full New Orleans data with eight lists, the lower threshold for the $p$-value yields a very different estimate, indeed one where the profile likelihood does not allow an upper 97.5\% confidence value, and a warning is generated by the routine within {\bf Rcapture}.  With a large number of lists and so many possible parameters to fit, it is not surprising that it should be inappropriate to use $p = 5\%$.  All the other estimates are similar to  estimates fitting main effects only, which are virtually unaffected by reducing to five lists.  

The Kosovo data yield a quite different result if interactions are allowed.  This is to be expected given the strong correlations evident in the data. 

\subsection{Identifiability and existence of estimates}
\label{subsec:identifiability}

The three data examples drawn from the study of human trafficking all give rise to contingency tables with some zero cell counts.   This raises issues discussed in generality by \cite{Fienberg2012}, and in our particular context by \cite{ChSiVi2019}.

One possibility is that there are no finite maximum likelihood estimates of all the parameters, but that the likelihood is maximized when one or more parameters tend to minus infinity.  This yields what \cite{Fienberg2012} term an {\em extended maximum likelihood estimate}, which gives a bona fide estimate, possibly zero, of each $\lambda_A$. This is handled within {\bf Rcapture}, somewhat unsatisfactorily, by returning large negative estimates for some of the $\beta_{ij}$.  These then give estimates for the resulting $\lambda_A$ which are very close to zero.  A consequence of this behaviour is that it is no longer possible to expand the log likelihood as a quadratic approximation around the maximum, and hence the standard likelihood theory, including the justification of information-based criteria, breaks down.   This breakdown is discussed further and illustrated in a small simulation example by \cite{ChSiVi2019}. 
Other aspects will be considered in 
Section \ref{subsec:uniformprior} below.  

There are two other estimability issues for maximum likelihood, neither of them addressed in {\bf Rcapture}.  
One is that the extended maximum likelihood estimate does not exist; consideration of a small artificial example in \cite{ChSiVi2019} shows that this may manifest itself as an infinite (or, numerically, very large) estimate of the dark figure.  
\cite{Fienberg2012a} show in a very general context that non-existence of the extended maximum likelihood estimate can be checked by solving a linear programming problem, set out  for our particular case in \cite{ChSiVi2019}.  The other possibility is that, although the likelihood 
can be maximized, the parameters that attain this maximum are unidentifiable; this can be checked by finding the rank of a particular model matrix.  

\cite{ChSiVi2019} derive an efficient algorithm that can demonstrate (without actually checking every single model) whether either check would be failed by any choice of the set of interaction parameters $\beta_{ij}$ to include in the model.   For each of the data sets considered in this paper, every possible model passes the checks. Thus it seems unlikely that the instabilities and bimodalities in the estimation displayed in Section \ref{subsec:modselapp}, and in some of the other methods discussed later in the paper, are due to the problems considered by \cite{Fienberg2012}.  

\section{Bayesian approaches}
\label{sec:bayes2}
Two rather different Bayesian approaches have been proposed or developed specifically for human rights data.  Their performance on our data sets is reviewed in this section.  Unfortunately, neither method escapes the instabilities already seen on the actual data sets.  

\subsection{Graphical models}
\label{sec:MadYork}
A graphical models method was developed by \citet{MadYor97} and implemented in the package {\bf dga} \citep{dgaR}.
This uses every decomposable graph model of dependencies between the various lists, and obtains the joint posterior probabilities of the models and the total population size.
The routine {\tt bma.cr} which carries out the analysis requires an array of possible values of the dark figure.  A reasonable standard range is from zero to 10 times the number of cases actually observed, but this will be discussed further below.  

The routine is only fully implemented for three, four and five lists, where the numbers of possible models are 8, 61 and 822 respectively. The combinatorial burden becomes excessive if six or more lists are used.   Therefore, the method is only applied on the five-list versions of the UK, Netherlands and New Orleans data, as well as on the Kosovo data and the four-list UK data.  

\begin{figure}
\centering
  \includegraphics[width=10cm]{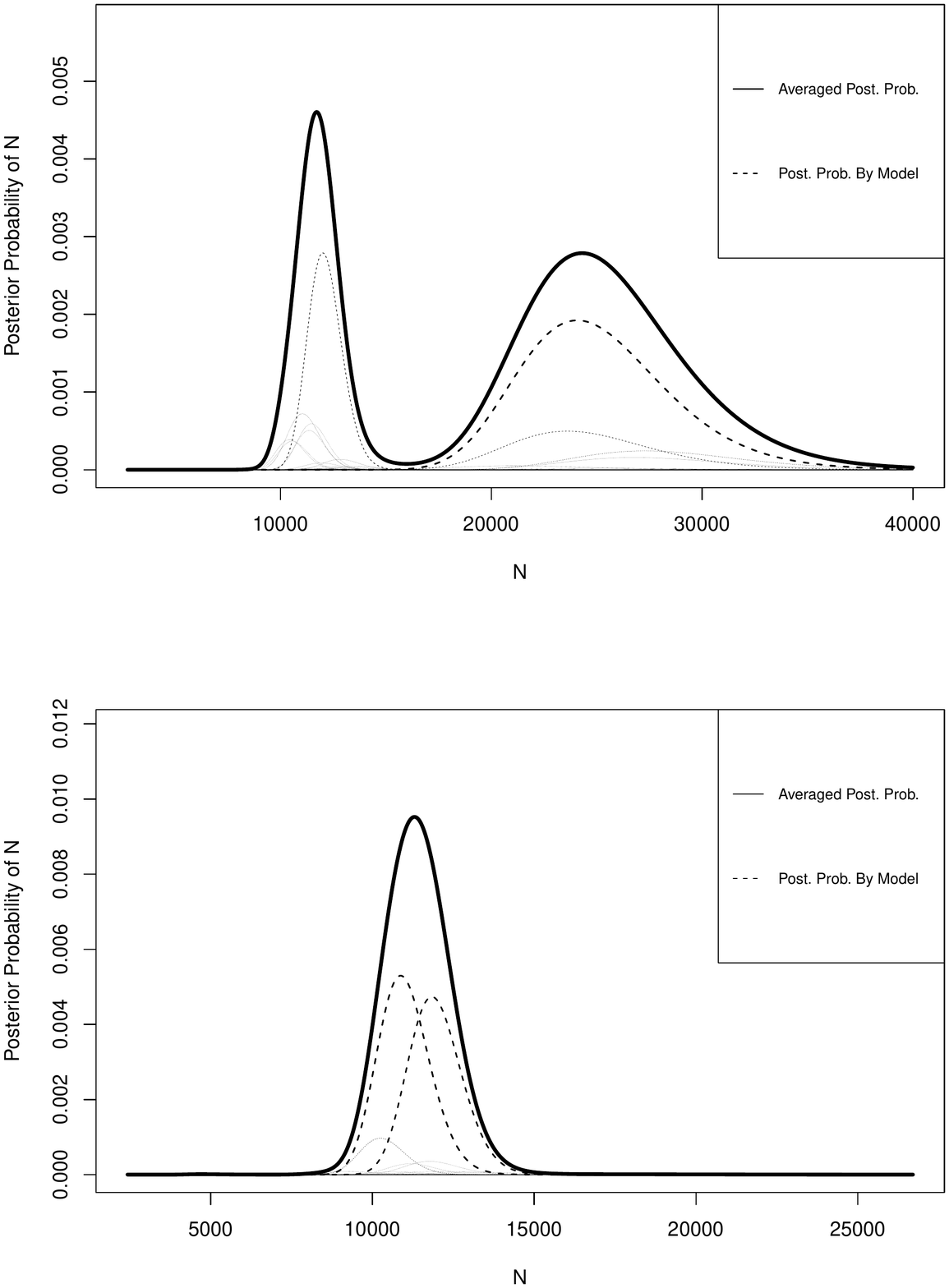}
  \caption{The posterior distribution of the total population size for the UK data, for the five-list and four-list data, using the method of Madigan and York}
  \label{fig:UKdatmadyorBoth}
\end{figure}

For the UK data, initial application of the method on the five-list data showed a strong bimodal distribution which extended beyond the standard range, and so the calculation was repeated with the range for the total population extended to 40000.  The results for both the five- and the four-list data are shown in Figure \ref{fig:UKdatmadyorBoth}.   The dotted curves show the joint posterior probabilities of particular values for the total population size and individual models.  The solid curve is the sum of the dotted curves, in other words the marginal posterior distribution of the total population size. There are 822 dotted curves in the upper figure in Figure 5, and 61 curves in the lower figure. Most of the models have posterior probability very close to zero for all values of the total population.  
The quantiles of the posterior distribution are given in Table \ref{tab:MadYork}, though of course in the case of the full five-list data these are not an adequate description of the bimodal distribution.  

\begin{table}
\caption{Quantiles of posterior distribution using the method of Madigan and York.  Where the value of the maximum population is given in the table, the range of possible population estimates is extended to that value beyond the default.
\label{tab:MadYork}}
\centering

\begin{tabular}{||lr|rrrrr|| }
\hline\hline
&Maximum&\multicolumn{5}{c||}{Quantiles of posterior}\\				
Data&population&2.5\%&10\%&50\%&90\%&97.5\%\\  \hline
\hline UK five lists&40&10.4&11.3&23.0&29.6&33.3 \\
UK four lists (GP omitted)&&9.5&10.1&11.4&12.7&13.5 \\
\hline Netherlands five lists&& 40.8 & 43.5 & 47.9 & 52.6 & 55.4 \\
Netherlands five lists &250&41.5 & 44.3 & 50.3 & 177.4 & 202.8 \\ \hline
\hline New Orleans five lists&&0.5 & 0.6 & 0.9 & 1.3 & 1.6 \\
\hline Kosovo && 9.8 & 10.8 & 12.6 & 14.9 & 16.5 \\ 
\hline\hline
\end{tabular}

\end{table}

\begin{figure}
\centering
  \includegraphics[width=10cm]{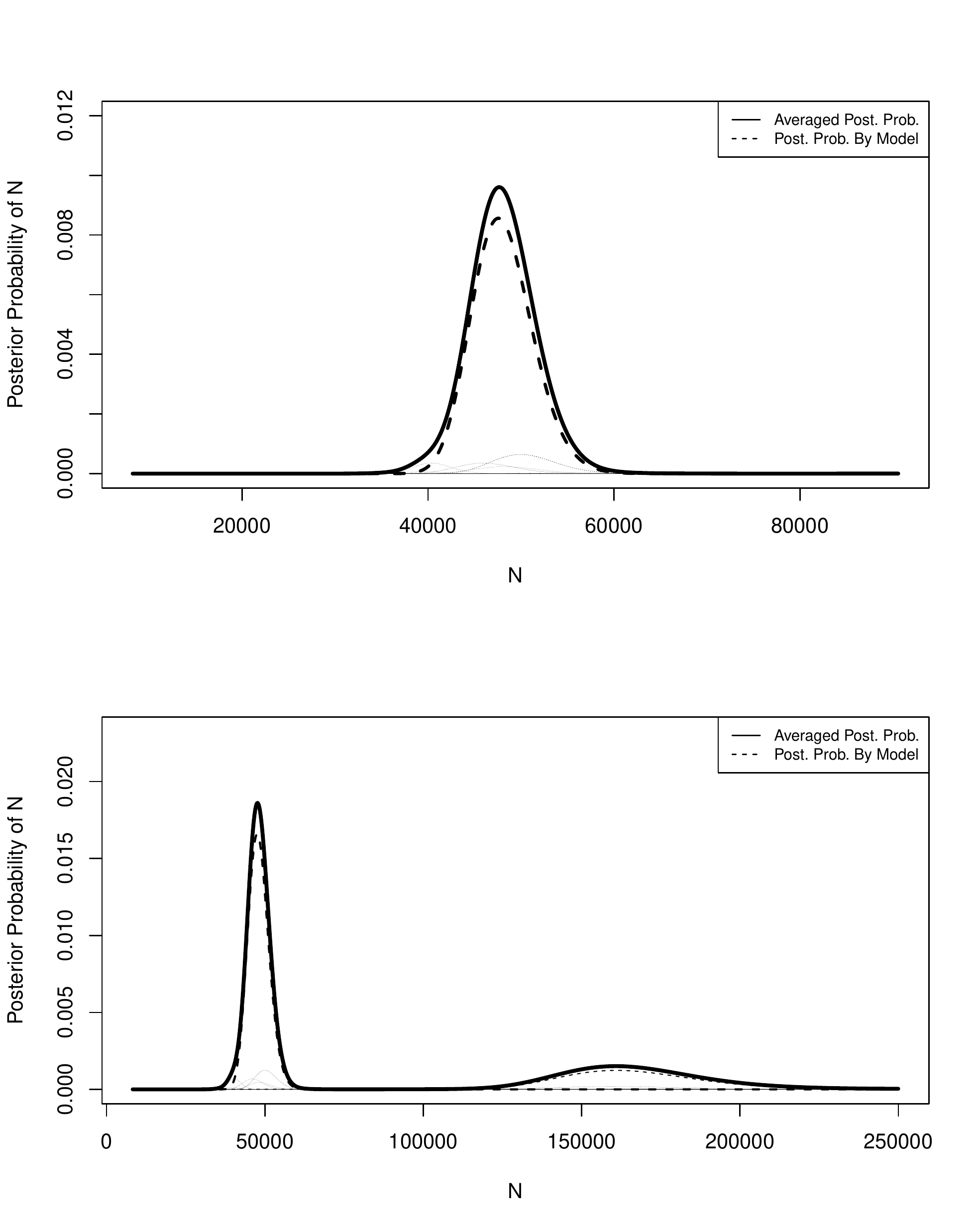}
  \caption{The posterior distribution of the total population size for the Netherlands data, using the method of Madigan and York.  Upper:  standard range of possible population size; lower: range extended to 250K.
  \label{fig:Nedmadyorcombined}}
\end{figure}

Now turn to the Netherlands data, where the number of observed cases is 8234.  The top panel of Figure \ref{fig:Nedmadyorcombined} shows the posterior when calculated on the range of up to ten times this figure for the dark figure.  In contrast with the UK data, there is no suggestion of any second mode within this range.  However, if the range is extended further, a noticeable mode appears, which has total posterior probability about 34\%.   The quantiles for the two estimates are given in Table \ref{tab:MadYork}.  

Results are also given in Table \ref{tab:MadYork} for the New Orleans and Kosovo data.   In these cases the posterior distribution is definitely concentrated within the standard range.  Interestingly, and in contrast with the other data considered, these two datasets illustrate two extremes of the method.  For the New Orleans data, the largest posterior probability of any of the possible models is about 0.05, so no model is dominant, while for the Kosovo data, one model has posterior probability nearly 0.99.  The corresponding probabilities (for the extended ranges) are 0.44 for the UK data and 0.67 for the Netherlands data.  



\clearpage

\subsection{Dirichlet process mixtures}

Another approach that has recently been proposed is a Bayesian latent class method \citep{Man16}.   This is implemented in the R package {\bf LCMCR} \citep{Man17R}.
It provides a Markov Chain Monte Carlo estimate of the population size.   In contrast with the method described in Section \ref{sec:MadYork} above, there is no restriction on the number of lists. The results for the various data sets are shown in Table \ref{tab:LCMCR}.

Because the output from the method is a Monte Carlo estimate, it is necessary to check whether there has been sufficient burn-in and also whether the output demonstrates sufficient mixing to be reliable.  
In order to ensure reproducibility the seed was set to 12345 rather than the default setting which yields different results each time.  To ensure better mixing than the default, the parameter {\tt thinning} was set to 100 and the {\tt burnin} value was set to 100K.

\begin{table}

\caption{Quantiles of posterior distribution using the Dirichlet process mixtures approach
\label{tab:LCMCR}}
\centering
\begin{tabular}{||l|rrrrr|| }
\hline\hline
&\multicolumn{5}{c|}{Quantiles of posterior}\\				
Data&2.5\%&10\%&50\%&90\%&97.5\%\\  
\hline \hline
UK six lists 	& 17.2 & 18.8 & 23.0 & 29.5 & 34.2 \\ 
UK five lists 	& 15.1 & 17.4 & 22.0 & 28.8 & 35.2 \\ 
UK four lists 	& 10.1 & 10.7 & 12.0 & 13.6 & 14.5 \\ \hline
Netherlands 	& 115.7 & 126.1 & 150.3 & 189.9 & 250.3 \\ 
Netherlands five lists & 43.0 & 44.7 & 49.1 & 54.2 & 58.3 \\ \hline
New Orleans & 0.5 & 0.6 & 0.7 & 0.9 & 1.1 \\ 
New Orleans five lists & 0.6 & 0.6 & 0.8 & 1.1 & 1.3 \\ \hline
Kosovo & 8.5 & 9.4 & 10.4 & 12.3 & 14.6 \\
\hline\hline
\end{tabular}\end{table}

Comparing the two Bayesian methods of this section is instructive.  For the UK data not omitting the GP list, the Dirichlet process approach essentially ignores the lower component of the posterior distribution found by the Madigan-York method and displayed in the upper panel of Figure \ref{fig:UKdatmadyorBoth}. 
Once GP is omitted, the two methods give very similar results.   For the Netherlands data, the Dirichlet approach homes in on the upper mode for the full data and the lower mode for the five-list case---the reverse of the behaviour of the AIC stepwise approach.

\section{The Bayesian/threshold approach}
\label{sec:bayes1}
\subsection{Defining the prior and thresholding the results}
In this section, we return to the Poisson log-linear model as specified in Section~\ref{subsec:poisspar}, and set out a Bayesian/threshold approach to fitting the model, dependent on two prior parameters, $\lambda$ and $\tau$.    The first step of the model is to specify a prior which does not constrain the intercept parameter or the main effects, but allows for the prior to shrink the interaction parameters towards zero.  In the second step, those interactions for which there is no strong evidence that they are not zero are dropped from the model, and the analysis repeated.   
The steps of the model are as follows.   
\begin{description}
\item[Step 1]
Use a prior model under which 
\begin{itemize}
\item
the parameters $\mu$, $\alpha_i$ and $\beta_{ij}$ for all $i$ and $j$ are independent
\item 
$\mu$ and the $\alpha_i$ have uniform (improper) prior on $( -\infty, \infty )$  
\item
the $\beta_{ij}$ have Gaussian prior with mean zero and variance $1/\lambda$ for $\lambda \ge 0$. If $\lambda=0$ this is interpreted as an improper uniform prior on $( -\infty, \infty )$.
\end{itemize}
In every case the R package {\bf MCMCpack} \citep{MQP11} and in particular the function {\tt MCMCpoisson} enable MCMC to be used to simulate from the posterior distribution.  The improper uniform prior is the default for parameters within {\tt MCMCpoisson}.  

\item[Step 2]
Constrain to zero those $\beta_{ij}$ for which the ratio of their posterior mean to their posterior standard deviation does not pass some threshold $\tau$, and repeat the MCMC analysis with these $\beta_{ij}$ omitted. 
\end{description}
One justification for the thresholding step is that it is an approximation to a prior for the interactions which is a mixture of an atom of probability at zero and some other distribution, a prior which in other contexts leads to a thresholding approach; see, for example, \cite{JoSi04}.  The exact implementation of such a prior is a topic for future research.  If $\tau=0$ then no thresholding is carried out.

In broad terms, a case is $\exp(\beta_{ij})$ times more or less likely to be on both the lists $i$ and $j$ than if occurrence on the lists is independent.  This interpretation makes it seem unlikely that values of $\beta_{ij}$ much outside the range $\pm 1$ should be contemplated, and so, if a Gaussian prior is used,  the precision parameter $\lambda$ might be chosen in the range 1 to 10.   

Turning to the thresholding parameter, two different approaches will be investigated.   The first is to take a `liberal' view, to include interactions where they are not clearly spurious; this would suggest using a threshold parameter of something like 2.   The other is to take a `parsimonious' view, using a much larger threshold, so that interaction parameters will only be included if there is very strong evidence that they are not zero.  For this approach we use a threshold of 5, admittedly chosen rather arbitrarily.

\subsection{Implementation issues}
\label{subsec:uniformprior}

There are two implementation issues taken into account in the package {\bf modslavmse} \citep{Sil18R}. Firstly, the  routine {\tt MCMCpoisson} in {\bf MCMCpack} does not appear to deal properly with the case where some of the parameters have an improper uniform distribution while others have finite variance, so if $\lambda>0$ the calling routine {\tt MCMCfit} in {\bf modslavmse} gives the intercept and main effects a prior with large finite variance $10^4$.   Note, in passing, that a proper Bayesian approach will avoid the issues considered in Section \ref{subsec:identifiability} because there will necessarily be a well-defined posterior distribution for the parameters. 

If an improper prior is used ($\lambda=0$) for the interaction parameters, then some care is needed. Consider the UK data as in Table \ref{tab:UKdat}.  No cases fall in both lists LA and GP, whether or not in combination with other lists.  If the improper uniform prior is used for the corresponding interaction parameter $\beta_{LA, GP}$, then we show that the posterior distribution of $\beta_{LA, GP}$ is concentrated at $-\infty$ and set out the way that the other parameters can be estimated by MCMC.  This is an instance where the maximum likelihood approach leads to an extended maximum likelihood estimate of the parameter;  see \cite{ChSiVi2019} for further discussion.  

In the general MSE model, suppose there is a pair of lists, without loss of generality lists 1 and 2, which contain no case in common.  Then $N_A=0$ for all combinations $A$ of lists containing both 1 and 2.  To find the posterior distribution of $\beta_{12}$, for each combination $B$ of lists, define
$$
C_B = \exp\left ( \mu + \sum_{i \in B}\alpha_i +  \sum_{\substack{i, j \in B, i<j\\ (i,j)\ne (1,2)}}  \beta_{ij}\right)
$$
It follows that 
\begin{displaymath}
\begin{array}{ll}
N_B \sim {\rm Poiss} ( C_B ) & {\rm if}   \{1,2 \} \not \subseteq B \\
N_B \sim {\rm Poiss} ( C_B \exp (\beta_{12})) & {\rm if}   \{1,2 \} \subseteq B 
\end{array}
\end{displaymath}
Because no cases are observed in the overlap of lists 1 and 2, we will have $N_B = 0$ for all $B \supseteq \{1,2\}$. So the conditional likelihood of $\beta_{12}$ given all the other parameters satisfies
\begin{eqnarray}
{\log L ( \beta_{12} |{\rm no~cases~in~common~between~1~and~2,~all~other~parameters} )}
\nonumber \\ \label{eq:condlik}
  =  -  \sum_{B \supseteq {\{1,2\}}}  C_B \exp (\beta_{12}) = - C \exp (\beta_{12}),
\end{eqnarray}
where $C > 0$ depends only on the parameters other than $\beta_{ij}$. Whatever the value of $C$, the log likelihood (\ref{eq:condlik}) is maximized as $\beta_{12} \rightarrow -\infty$.

The posterior density of $\beta_{12}$ is proportional to $\exp ( -C e^\beta )$.  Although this appears at first sight to be an improper distribution, this function has the properties that, for all $y$,  
$$
\int_{-\infty}^{y} \exp ( -C e^\beta ) d \beta = \infty {\rm ~and} \int_y ^{\infty} \exp ( -C e^\beta ) d \beta < \infty
$$
so that $P(\beta_{12} > y )/P(\beta_{12} \le y) = 0)$.  
This corresponds to the distribution where $\beta_{12} = -\infty$ with probability 1.  Since this is true conditional on all the other parameters whatever their values, the unconditional posterior distribution is the same.  Hence the posterior distribution of the Poisson parameter for every $B$ that includes lists 1 and 2 is an atom of probability at 0.  Given the value $-\infty$ for $\beta_{12}$, the distribution of every $N_B$ for each $B \supseteq \{1,2\}$ is then Poisson with parameter 0, in other words the constant value 0, regardless of the other parameters, while for all other $B$, $N_B \sim {\rm Poiss}(\lambda_B)$ with $\lambda_B$ defined as in equation (\ref{eq:poisspar}) above.  So, as asserted above, the likelihood of all the other parameters conditional on $\beta_{12}= -\infty$ is then obtained by simply omitting all combinations of lists which contain 1 and 2. 

Returning to the UK data example, where there are 6 lists and hence 63 observable combinations $B$, we omit the 16 $N_B$ for which $B$ included both lists LA and GP, leaving 47 observations from which to estimate the remaining 21 parameters.   In fact there is a second pair of lists for which there is no overlap at all, namely LA and NCA, and by the same argument the parameter $\beta_{LA, NCA}$ is also estimated to be $-\infty$ with probability one.  Removing (from the 47 remaining combinations) all combinations of lists containing both LA and NCA leaves 39 observations from which to apply the MCMC approach to the remaining 20 parameters.    Within the package {\bf modslavmse}, the routine {\tt removeemptyoverlaps}, which is called from {\tt MCMCfit}, produces the relevant data matrix and also a list of those interaction parameters that take the value $-\infty$ in the posterior. 

\subsection{Results}

In this section, the results for the three examples are presented, exploring the effects of using various priors and various thresholds. 

The results for the UK data are given in Tables \ref{tab:UKdat1results}, \ref{tab:UKdat2results} and \ref{tab:UKdat3results}.  The first line in each table shows the result of fitting the main effects only, with no $\beta_{ij}$ considered.  
Once interactions are considered, the results are not enormously sensitive to the prior, especially if a non-zero threshold is used.  If there is no thresholding, so that all interactions are included within the model, then the posterior credible intervals are much larger, but the central estimate is similar.  

Computationally, the uniform improper prior is the fastest, though the possibly more plausible prior with variance 1 gives much the same results.  A model with every possible interaction is more complicated than the amount of data can bear, and the thresholding at a threshold of 2 is a liberal approach which nevertheless eliminates extraneous complication.  This would tend to suggest a point estimate of about 12.3K for the overall prevalence, with an 80\% credible interval (rounding to the nearest 500) of about 11.5K to 13.5K and a 95\% credible interval, in round terms, of 11K to 14K.  This is about a thousand more than the confidence interval obtained from the fixed model, but this is possibly because of the model averaging taking some note of the second group of models exemplified by the model chosen by the BIC criterion.    Increasing the threshold to 5 makes little difference for the full data, and, as we see below, homes in on just a single interaction among the lists.  

\begin{table}
\caption{Quantiles of posterior distribution of the total population size, including both the observed data and the dark figure, UK data with six lists
\label{tab:UKdat1results}}
\centering
\begin{tabular}{||lr|rrrrr|| }
\hline\hline
&&\multicolumn{5}{c||}{Quantiles of posterior}\\				
Prior&Threshold&2.5\%&10\%&50\%&90\%&97.5\%\\ \hline 
\multicolumn{2}{||c|}{Main effects only}&
										11.0 	& 11.4 	& 12.2 	& 13.1 	& 13.6\\ 
\hline Uniform 	& 0 	& 6.2 		& 7.5 		& 10.9 	& 15.3 	& 18.4\\ 
Variance 10 	   & 0 	& 7.0 		& 8.5 		& 10.3 	& 15.4 	& 20.4\\ 
Variance 1 	      & 0 	& 8.5 		& 9.3 		& 13.0 	& 15.5 	& 17.1\\ 
Variance 0.1 	   & 0 	& 10.2 	& 10.3 	& 11.9 	& 13.5 	& 14.3\\ 
\hline Uniform 	& 2 	& 10.7 	& 11.2 	& 12.2 	& 13.6 	& 14.4\\ 
Variance 10 	   & 2 	& 10.9 	& 11.4 	& 12.3 	& 13.5 	& 14.3\\ 
Variance 1     	& 2 	& 10.9 	& 11.3 	& 12.3 	& 13.6 	& 14.4\\ 
Variance 0.1 		& 2 	& 10.8 	& 11.2 	& 12.0 	& 13.0 	& 13.3\\ 
\hline Uniform 	& 5 	& 10.9 	& 11.2 	& 12.1 	& 13.0 	& 13.6\\ 
Variance 10 		& 5 	& 11.1 	& 11.5 	& 12.3 	& 13.1 	& 13.8\\ 
Variance 1 			& 5 	& 11.4 	& 11.9 	& 12.7 	& 13.8 	& 14.2\\ 
Variance 0.1 		& 5 	& 11.1 	& 11.5 	& 12.3 	& 13.1 	& 13.8\\ 
\hline \hline
\end{tabular} 
\end{table}

\begin{table}
 \caption{Quantiles of posterior distribution of the total population size, including both the observed data and the dark figure, UK data with five lists \label{tab:UKdat2results}}
\centering

\begin{tabular}{||lr|rrrrr|| }
\hline\hline
&&\multicolumn{5}{c||}{Quantiles of posterior}\\				
Prior&Threshold&2.5\%&10\%&50\%&90\%&97.5\%\\ \hline 
\multicolumn{2}{||c|}{Main effects only}&
									12.0	& 12.5	& 13.5	& 14.6	& 15.3\\ 
\hline Uniform	& 0	& 5.9		& 7.7		& 11.1	& 18.4	& 22.2\\ 
Variance 10		& 0	& 6.2		& 7.8		& 13.0	& 19.7	& 24.5\\ 
Variance 1			& 0	& 9.4		& 10.8	& 13.8	& 19.1	& 23.0\\ 
Variance 0.1		& 0	& 11.4	& 12.3	& 14.5	& 17.2	& 18.5\\ 
\hline Uniform	& 2	& 10.7	& 11.1	& 12.2	& 13.3	& 13.9\\ 
Variance 10		& 2	& 11.6	& 12.0	& 13.1	& 14.1	& 14.7\\ 
Variance 1			& 2	& 11.7	& 12.1	& 13.2	& 14.3	& 15.1\\ 
Variance 0.1		& 2	& 12.2	& 12.8	& 14.1	& 15.4	& 16.2\\ 
\hline Uniform	& 5	& 12.0	& 12.4	& 13.3	& 14.4	& 15.1\\ 
Variance 10		& 5	& 12.0	& 12.5	& 13.5	& 14.6	& 15.3\\ 
Variance 1			& 5	& 12.6	& 13.1	& 14.1	& 15.3	& 16.0\\ 
Variance 0.1		& 5	& 12.0	& 12.5	& 13.5	& 14.6	& 15.3\\ 

\hline \hline
\end{tabular}\end{table}

\begin{table}
\caption{Quantiles of posterior distribution of the total population size, including both the observed data and the dark figure, UK data with four lists (five-list data with GP omitted) \label{tab:UKdat3results}}
\centering

\begin{tabular}{||lr|rrrrr|| }
\hline\hline
&&\multicolumn{5}{c||}{Quantiles of posterior}\\				
Prior&Threshold&2.5\%&10\%&50\%&90\%&97.5\%\\ \hline 
\multicolumn{2}{||c|}{Main effects only} &
										9.5	& 9.9	& 10.7	& 11.6	& 12.1\\ 
\hline Uniform	& 0	& 6.4	& 8.1	& 12.1	& 18.5	& 23.2\\ 
Variance 10		& 0	& 6.1	& 7.7	& 11.3	& 16.7	& 21.5\\ 
Variance 1			& 0	& 6.7	& 7.6	& 10.1	& 14.3	& 17.8\\ 
Variance 0.1		& 0	& 7.5	& 8.2	& 9.6	& 11.4	& 12.6\\ 
\hline Uniform	& 2	& 10.5	& 11.0	& 12.0	& 13.1	& 13.7\\ 
Variance 10		& 2	& 10.6	& 11.0	& 12.0	& 13.0	& 13.7\\ 
Variance 1			& 2	& 10.4	& 10.9	& 11.8	& 12.9	& 13.7\\ 
Variance 0.1		& 2	& 9.3	& 9.7	& 10.6	& 11.5	& 11.9\\ 
\hline Uniform	& 5	& 9.5	& 9.9	& 10.7	& 11.6	& 12.1\\ 
Variance 10		& 5	& 9.5	& 9.9	& 10.7	& 11.6	& 12.1\\ 
Variance 1			& 5	& 9.5	& 9.9	& 10.7	& 11.6	& 12.1\\ 
Variance 0.1		& 5	& 9.5	& 9.9	& 10.7	& 11.6	& 12.1\\ 

\hline \hline

\end{tabular} 
\end{table}

\begin{table}
\caption{Quantiles of posterior distribution of the total population size, including both the observed data and the dark figure, Netherlands data.  
\label{tab:Nedresults}}
\centering

\begin{tabular}{||lr|rrrrr|| }
\hline\hline
&&\multicolumn{5}{c||}{Quantiles of posterior}\\				
Prior&Threshold&2.5\%&10\%&50\%&90\%&97.5\%\\  \hline
\multicolumn{2}{||c|}{Main effects only}&
48.7	& 49.9	& 52.6	& 55.9	& 57.9\\ 
\hline Uniform	& 0	& 31.0	& 36.0	& 50.9	& 65.3	& 72.4\\ 
Variance 10		& 0	& 35.7	& 36.6	& 52.0	& 73.1	& 83.5\\ 
Variance 1			& 0	& 46.5	& 52.0	& 69.3	& 74.5	& 81.5\\ 
Variance 0.1		& 0	& 49.1	& 53.6	& 60.9	& 67.7	& 71.5\\ 
\hline Uniform	& 2	& 42.4	& 44.4	& 47.6	& 52.2	& 54.7\\ 
Variance 10		& 2	& 43.5	& 44.2	& 47.3	& 52.2	& 53.5\\ 
Variance 1			& 2	& 60.6	& 64.0	& 73.0	& 85.6	& 93.0\\ 
Variance 0.1		& 2	& 56.9	& 59.2	& 66.2	& 74.1	& 78.7\\ 
\hline Uniform	& 5	& 51.3	& 52.9	& 56.1	& 59.1	& 60.9\\ 
Variance 10		& 5	& 54.7	& 56.0	& 59.5	& 63.2	& 65.4\\ 
Variance 1			& 5	& 54.6	& 56.2	& 59.4	& 62.9	& 65.2\\ 
Variance 0.1		& 5	& 61.0	& 63.4	& 68.2	& 73.3	& 75.9\\ 
\hline\hline
\end{tabular}
\end{table}

\begin{table}
\caption{Quantiles of posterior distribution of the total population size, including both the observed data and the dark figure, New Orleans data consolidated into five lists
\label{tab:NewOrlresults}}
\centering

\begin{tabular}{||lr|rrrrr|| }
\hline\hline
&&\multicolumn{5}{c||}{Quantiles of posterior}\\				
Prior&Threshold&2.5\%&10\%&50\%&90\%&97.5\%\\ \hline 
\multicolumn{2}{||c|}{Main effects only}&
0.7 & 0.8 & 1.1 & 1.5 & 1.9 \\ 
\hline Uniform & 0 & 0.4 & 0.8 & 4.1 & 17.4 & 38.7 \\ 
Variance 10 	& 0 & 0.6 & 1.2 & 2.8 & 10.5 & 24.2 \\ 
Variance 1 & 0 & 0.6 & 0.8 & 1.3 & 2.4 & 2.9 \\ 
Variance 0.1 & 0 & 0.6 & 0.8 & 1.0 & 1.5 & 1.8 \\ 
\hline Uniform & 2 & 0.6 & 0.6 & 0.8 & 1.2 & 1.3 \\
 Variance 10 & 2 & 0.8 & 0.9 & 1.2 & 1.8 & 2.3 \\ 
 Variance 1 & 2 & 0.7 & 0.8 & 1.1 & 1.5 & 1.9 \\ 
 Variance 0.1 & 2 & 0.7 & 0.8 & 1.1 & 1.5 & 1.9 \\ 
 \hline Uniform & 5 & 0.6 & 0.6 & 0.8 & 1.2 & 1.3 \\ 
 Variance 10 & 5 & 0.7 & 0.8 & 1.1 & 1.5 & 1.9 \\ 
 Variance 1 & 5 & 0.7 & 0.8 & 1.1 & 1.5 & 1.9 \\ 
 Variance 0.1 & 5 & 0.7 & 0.8 & 1.1 & 1.5 & 1.9 \\
\hline \hline
\end{tabular}
\end{table}

Now turn to the Netherlands data, the results for which are shown in Table~\ref{tab:Nedresults}.  Again, and not surprisingly, if all interactions are considered in the model then the posterior intervals are much wider.  However, if the thresholding procedure is used to restrict attention to a smaller number of interactions, then the width of the intervals is not dramatically different from the main effects model.  
Threshold 2 with variance 1 appears to be an exception; however, examination of the results shows that only 5 of the 15 two-factor interactions are thresholded out. As a check, the method was run on the five-list version of the data, with the two smallest lists consolidated.  The results for the five-list data were, in general, slightly lower for thresholds 0 and 2 and slightly higher for threshold 5.  The only substantially different case was variance 1 threshold 2, where the five-list data results are about 70\% of the result for the six-list data. 

The New Orleans data are a smaller set of observations, and also consist of eight lists with none of the overlap sets containing more than two cases.  Therefore it does not seem appropriate to use more than the main effects model and that approach was adopted in the original analysis \citep{BaMuSi18}.  However it is of interest to see what would happen if we use the Bayesian approach allowing for interactions. Some trials suggest that if the full 8 list data are used, the MCMC algorithm requires both a long burn-in period and then a long run, and possibly other adjustments to the control parameters, to give reasonable mixing in the posterior realisations.  For simplicity, therefore, we analyse the five-list version, and the results are given in Table~\ref{tab:NewOrlresults}.  The variance 1 threshold 2 model (and indeed some of the other models) give results identical to the main effects only model, and closer examination of the estimates within the package shows that the thresholding step in fact removes all the interactions leaving main effects only. 

However, the uniform prior, even with strong thresholding, gives different estimates.  To understand why, note that there are 10 two-factor interactions $\beta_{ij}$ between the five lists.  In three of these cases, the observed overlap between lists $i$ and $j$ is zero, and so the corresponding $\beta_{ij}$ is estimated as $-\infty$ regardless of the thresholding.   Even with a moderate threshold all the other interactions are thresholded out, but the model is fitted not just based on the main effects only but with three of the interactions included and estimated to $-\infty$.   If the original eight-list data are considered, then the effect is much stronger, with 18 of the 28 possible interaction parameters estimated as $-\infty$. 


\begin{table}
\caption{Quantiles of posterior distribution of the total population size, including both the observed data and the dark figure, Kosovo data
\label{tab:Kosovoresults}}
\centering

\begin{tabular}{||lr|rrrrr|| }
\hline\hline
&&\multicolumn{5}{c||}{Quantiles of posterior}\\ 
Prior&Threshold&2.5\%&10\%&50\%&90\%&97.5\%\\ \hline 
\multicolumn{2}{||c|}{Main effects only}
&  7.2 &  7.2 &  7.4 &  7.6 &  7.7\\
\hline \hline  
Uniform &  0 &  12.5 &  13.1 &  14.4 &  15.9 &  16.7\\
 variance 10 &  0 &  12.5 &  13.1 &  14.3 &  15.9 &  16.8\\
 variance 1 &  0 &  12.3 &  12.9 &  14.1 &  15.6 &  16.4\\
 variance 0.1 &  0 &  10.7 &  11.1 &  12.1 &  13.2 &  13.8\\
 \hline \hline Uniform &  2 &  12.6 &  13.1 &  14.3 &  15.7 &  16.3\\
 variance 10 &  2 &  12.6 &  13.1 &  14.2 &  15.5 &  16.4\\
 variance 1 &  2 &  12.3 &  12.9 &  14.0 &  15.2 &  16.1\\
 variance 0.1 &  2 &  10.9 &  11.2 &  12.1 &  13.1 &  13.6\\
 \hline \hline Uniform &  5 &  12.6 &  13.1 &  14.3 &  15.7 &  16.3\\
 variance 10 &  5 &  12.6 &  13.1 &  14.2 &  15.5 &  16.4\\
 variance 1 &  5 &  12.3 &  12.9 &  14.0 &  15.2 &  16.1\\
 variance 0.1 &  5 &  10.9 &  11.2 &  12.1 &  13.1 &  13.6\\
\hline \hline
\end{tabular}

\end{table}

The Kosovo data are unusual in that all the models allowing for interactions give broadly similar results; see Table~\ref{tab:Kosovoresults}.  The thresholding has little or no effect, even at a threshold of 5, because most of the interactions are very strong.   

\subsection{Choosing the threshold for interactions}

The Bayesian approach avoids the necessity of choosing a particular model, but it still contains tuneable prior parameters.  The implausibility of very large positive or negative values for the interaction parameters suggests that a prior variance of 1 is a reasonable choice.  The standard MCMC software does not allow for the mixed model with an atom of probability at zero for the parameters, a topic for future research,  but the thresholding approach gives a simple alternative.  

\begin{table}
\caption{Interactions included in the variance 1, threshold 2 model.  For threshold 5, only the effects shown in bold survive the thresholding step. For the four-list UK data, the effect LA:NG does survive up to thresholds of about 3.5.
\label{tab:interactions}}
\centering

\begin{tabular}{||l|l|l|| }
\hline\hline
Data set &Lists& Interactions included \\
\hline 
UK  & 6 & {\bf LA:NG}, LA:PF, NG:GP, PF:GP, PF:NCA, GO:GP \\
UK & 5 & {\bf LA:NG}, LA:PFNCA, NG:GP, PFNCA:GP \\
UK excluding GP  & 4& LA:NG, LA:PFNCA \\ \hline
Netherlands & 6 & I:K, I:Z, K:O, K:P, K:R, K:Z, O:P, {\bf O:Z},  P:R, P:Z \\ \hline
New Orleans & 8 & no interactions at either threshold \\ \hline
Kosovo & 4 & {\bf all except ABA:HRW} \\
\hline \hline
\end{tabular}

 \end{table}
 
In Table~\ref{tab:interactions} we see the interactions that exceed the threshold at the first stage for both thresholds considered.  The three results for the UK data are entirely consistent with one another, given that the second data set is obtained by combining the PF and NCA lists and the third by omitting GP.  
For the Netherlands data, 10 of the possible 15 interactions survive a threshold of 2, and the results obtained are somewhat anomalous, both when compared with those for other parameter values and when compared with the effect of combining the two smallest lists. For threshold 5, the method picks out the LA:NG interaction only for the UK data and the O:Z interaction for the Netherlands data, with the same results in both cases if the two smallest lists are consolidated.  Leaving aside prevalence estimation as such, an advantage of the more parsimonious approach is that it focuses in on those pairs where there is a very clear interaction, giving pointers as to where to particularly look to gain a greater understanding of what is going on.   However, it is intuitively clear in the modern slavery case that correlations between lists are not at all surprising, and the results demonstrated in Table~\ref{tab:interactions} suggest that the less restrictive threshold 2 is probably to be preferred at least as a starting point.   Interestingly, reducing the threshold in the Netherlands data to 4.5 yields a similar result to threshold 2.  

For the New Orleans data, any reasonable level of thresholding leads back to the fitting of main effects only, which is probably the most realistic model given the number of lists and the numbers of cases in the various overlaps.  On the other hand, for the Kosovo data, only one of the interaction effects is thresholded out, even at the high threshold.  This is not surprising since the data clearly demonstrate strong inter-list correlations, and it is very reassuring that even the high threshold adapts well to data of this kind.  

Overall, consideration of these examples suggests that the Bayesian/threshold model with variance 1 and threshold 2 adapts reasonably well to the characteristics of different data sets, although it is advisable not to apply the method completely blindly.

\section{Conclusions}
\label{sec:conclusions}
Estimating and keeping track of the numbers of victims is a crucial component of the fight against modern slavery.  If multiple systems estimation is to be used as one of the standard methods, then the stability and robustness of point and interval estimation is an important consideration.  The most stable method would, of course, be to ignore the possibility of interactions and simply fit main effects, but the Kosovo example shows that this would clearly be inadequate in some practical cases.  It would also fail to take account of the correlations which are not unexpected between the lists obtained in the modern slavery context.   

The Bayesian approach of this paper, with a threshold of 2 and a prior variance of 1 for the interaction parameters, is at least a candidate.  On the data sets considered, it gives results which are stable and robust when smaller lists are combined, and it automatically rules out implausible secondary estimates which are almost certainly spurious.  If it is desirable to obtain parsimonious explanations in cases where there may be interactions of particular interest, then the threshold can if necessary be increased.  The approach adapts well between data such as the Kosovo data, with strong dependencies between lists, and those situations where few, if any, interactions are clearly present in the data.  

One contrast between the modern slavery data sets and the Kosovo data is that the modern slavery data are much sparser, in that not every combination of lists is observed at all.   This is not a reflection of the data quality, but is intrinsic to the field.  In modern slavery, we will often wish to quantify the number of victims in a fairly constrained geographical area over a reasonably short time period, and so the total population size may be quite small, as in the Greater New Orleans example.  Even when we consider larger data sets, such as the Netherlands data, the number of cases actually observed may only be a relatively small proportion of the total population.  Sparse data, and lists that do not overlap at all, are the norm rather than the exception, and methods need to take account of that.  Of course, it is to be hoped that as public and political consciousness about modern slavery increases, a larger proportion of cases will actually come to light, but this is likely to be a long process.  Further recent work on this aspect by the author and colleagues is reported in \cite{ChSiVi2019}.
It should also be noted that when multiple systems estimation is used for a census of an animal or an easy-to-count human population, attempts can be made to design the surveys or captures to be independent of one another and also to be large enough to avoid the sparsity issues raised by the modern slavery data sets;  however, in most human rights contexts, there is no such control over the way that lists arise.

The availability of real data has been an important contribution to the study carried out in this paper, because data on modern slavery and human trafficking will have specific characteristics which need to be taken into account.  It is to be hoped that more data sets will be put into the public domain, of course in formats that preserve the privacy of individuals and do not hamper the primary task of rescuing and supporting victims, bringing perpetrators to justice, and discouraging modern slavery in the future.    

Multiple systems estimation is not a panacea, but part of the quest for better information and understanding.  A key topic for discussion and for future research is how we can build on a whole range of different information and methods to gain a deeper understanding of modern slavery.  For example, a promising development is the typology developed by \cite{Cooper2017} and the associated case file coding template.  More widely, the important role of research in fighting modern slavery is underlined by the research priorities set out in the Government's 2018 Annual Report on Modern Slavery \citep{HMGov2018}.   A broad discussion of the actual and potential modes of measurement, and how these fit in to the legal, definitional and historical background of modern slavery, is given by \cite{Landman2019}.   

There are several avenues for future research on the multiple systems methodology set out in this paper.  For example, how can it be developed to handle concomitant information and segmentation of populations?  What is the best approach when the aim is to discern whether the overall level is different between two time points or between two different sectors or geographical areas?  Can the approach be easily extended to the case of fuzzy matching, where it is not quite clear whether cases on different lists are or are not the same?  Perhaps most importantly, are there particular patterns in data sets drawn in the context of modern slavery and human trafficking, and can these, as well as the prevalence estimates themselves, contribute to a deeper understanding of the problem itself?  

\section*{Acknowledgements}

The author gratefully acknowledges correspondence and other help from Kevin Bales, Patrick Ball, Peter van der Heijden, Ella Kaye and Daniel Manrique-Vallier, and the very helpful comments of the referees. This work was supported by the Arts and Humanities Research Council and the Economic and Social Research Council grant [number ES/P001491/1] Modern Slavery: Meaning and Measurement (PaCCS Transnational Organised Crime, University of Nottingham, 2016--18).

\bibliography{modslav}

\end{document}